\pdfoutput=1
\documentclass[runningheads]{llncs}

\usepackage[T1]{fontenc}
\usepackage{graphicx}

\usepackage{acro}        
\usepackage{array}
\usepackage{tikz}        
\usepackage{xcolor}      
\usepackage{pifont}      
\usepackage{stackengine} 
\usepackage{rotating}
\usepackage{enumitem}
\usepackage{wrapfig}
\usepackage{tabularx}

\setlist{noitemsep, topsep=2pt, partopsep=0pt, parsep=0pt}
\usepackage{hyperref}

\def\vordoi{}
\def\vorvolume{Secure IT Systems: 31st Nordic Conference, NordSec 2026, Proceedings, Springer Nature}
\usepackage{color}

\newcommand{\crossedcmark}{\stackinset{c}{}{c}{}{\xmark}{\cmark}}
\newcommand{\cmark}{\ding{51}}
\newcommand{\xmark}{\ding{55}}

\DeclareAcronym{imsi}{
  short        = IMSI ,
  short-plural = IMSIs ,
  long         = International Mobile Subscriber Identity ,
  long-plural  = International Mobile Subscriber Identities
}
\DeclareAcronym{arfcn}{
  short        = ARFCN ,
  long         = Absolute Radio Frequency Channel Number
}
\DeclareAcronym{sa}{
  short        = SA ,
  long         = Standalone
}
\DeclareAcronym{pdcp}{
  short        = PDCP ,
  long         = Packet Data Convergence Protocol
}
\DeclareAcronym{rlc}{
  short        = RLC ,
  long         = Radio Link Control
}
\DeclareAcronym{mn}{
  short = MN,
  long  = Master Node
}
\DeclareAcronym{oem}{
  short = OEM,
  long  = Original Equipment Manufacturer
}
\DeclareAcronym{soc}{
  short = SoC,
  long = System-on-Chip
}
\DeclareAcronym{nas}{  short = NAS ,  long  = Non-Access-Stratum}
\DeclareAcronym{rrc}{  short = RRC ,  long  = Radio Resource Control}
\DeclareAcronym{dos}{  short = DoS ,  long  = Denial-of-Service}
\DeclareAcronym{ran}{  short = RAN ,  long  = Radio Access Network}
\DeclareAcronym{ue}{  short = UE ,  long  = User Equipment ,  short-plural = s,  long-plural  = s
}
\DeclareAcronym{emm}{short=EMM, long=EPS Mobility Management}
\DeclareAcronym{fbs}{short=FBS, long=Fake Base Station}
\DeclareAcronym{nr}{short=NR, long=New Radio}
\DeclareAcronym{lte}{short=LTE, long=Long Term Evolution}
\DeclareAcronym{eutra}{short=E-UTRA, long=Evolved Universal Terrestrial Radio Access}
\DeclareAcronym{enb}{short=eNB, long=Evolved Node B}
\DeclareAcronym{gnb}{short=gNB, long=Next Generation Node B}
\DeclareAcronym{cn}{short=CN, long=Core Network}
\DeclareAcronym{epc}{short=EPC, long=Evolved Packet Core}
\DeclareAcronym{mme}{short=MME, long=Mobility Management Entity}
\DeclareAcronym{sgw}{short=SGW, long=Serving Gateway}
\DeclareAcronym{pgw}{short=PGW, long=Packet Gateway}
\DeclareAcronym{hss}{short=HSS, long=Home Subscriber Server}
\DeclareAcronym{5gc}{short=5GC, long=5G Core}
\DeclareAcronym{amf}{short=AMF, long=Access and Mobility Management Function}
\DeclareAcronym{smf}{short=SMF, long=Session Management Function}
\DeclareAcronym{upf}{short=UPF, long=User Plane Function}
\DeclareAcronym{udm}{short=UDM, long=Unified Data Management}
\DeclareAcronym{udr}{short=UDR, long=Unified Data Repository}
\DeclareAcronym{mib}{short=MIB, long=Master Information Block}
\DeclareAcronym{sib}{short=SIB, long=System Information Block}
\DeclareAcronym{prach}{short=PRACH, long=Physical Random Access Channel}
\DeclareAcronym{rar}{short=RAR, long=Random Access Response}
\DeclareAcronym{ul}{short=UL, long=Uplink}
\DeclareAcronym{crnti}{short=C-RNTI, long=Cell Radio Network Temporary Identifier}
\DeclareAcronym{s1ap}{short=S1AP, long=S1 Application Protocol}
\DeclareAcronym{ngap}{short=NGAP, long=Next Generation Application Protocol}
\DeclareAcronym{3gpp}{short=3GPP, long=3rd Generation Partnership Project}
\DeclareAcronym{as}{short=AS, long=Access Stratum}
\DeclareAcronym{plmn}{short=PLMN, long=Public Land Mobile Network}
\DeclareAcronym{tau}{short=TAU, long=Tracking Area Update}
\DeclareAcronym{tai}{short=TAI, long=Tracking Area Identity}
\DeclareAcronym{tac}{short=TAC, long=Tracking Area Code}
\DeclareAcronym{imei}{short=IMEI, long=International Mobile Equipment Identity}
\DeclareAcronym{rat}{short=RAT, long=Radio Access Technology}
\DeclareAcronym{ta}{short=TA, long=Tracking Area}
\DeclareAcronym{pci}{short=PCI, long=Physical Cell Identity}
\DeclareAcronym{earfcn}{short=EARFCN, long=E-UTRA Absolute Radio Frequency Channel Number}
\DeclareAcronym{sdr}{short=SDR, long=Software Defined Radio}
\DeclareAcronym{rf}{short=RF, long=Radio Frequency}
\DeclareAcronym{usrp}{short=USRP, long=Universal Software Radio Peripheral}
\DeclareAcronym{mcc}{short=MCC, long=Mobile Country Code}
\DeclareAcronym{mnc}{short=MNC, long=Mobile Network Code}
\DeclareAcronym{pdu}{short=PDU, long=Packet Data Unit}
\DeclareAcronym{ccch}{short=CCCH, long=Common Control Channel}
\DeclareAcronym{bcch}{short=BCCH, long=Broadcast Control Channel}
\DeclareAcronym{srb}{short=SRB, long=Signaling Radio Bearer}
\DeclareAcronym{drb}{short=DRB, long=Data Radio Bearer}
\DeclareAcronym{mcg}{short=MCG, long=Master Cell Group}
\DeclareAcronym{rlf}{short=RLF, long=Radio Link Failure}
\DeclareAcronym{endc}{short=EN-DC, long=E-UTRA-NR Dual Connectivity}
\DeclareAcronym{sn}{short=SN, long=Secondary Node}
\DeclareAcronym{scg}{short=SCG, long=Secondary Cell Group}
\DeclareAcronym{geran}{short=GERAN, long=GSM EDGE Radio Access Network}
\DeclareAcronym{utra}{short=UTRA, long=UMTS Terrestrial Radio Access}
\DeclareAcronym{asn1}{short=ASN.1, long=Abstract Syntax Notation One}
\DeclareAcronym{pki}{short=PKI, long=Public-Key Infrastructure}
\DeclareAcronym{hibs}{short=HIBS, long=Hierarchical Identity-Based Signature}
\DeclareAcronym{qos}{short=QoS, long=Quality of Service}
\DeclareAcronym{mac}{short=MAC, long=Message Authentication Code}

\DeclareAcronym{sim}{short=SIM, long=Subscriber Identity Module}
\DeclareAcronym{cots}{short=COTS, long=Commercial off-the-shelf}
\DeclareAcronym{bs}{short=BS, long=Base Station}
\DeclareAcronym{iot}{short=IoT, long=Internet of Things}
\DeclareAcronym{oran}{short=O-RAN, long=Open RAN}

\begin{document}

\title{RRC Ping-Pong: Pre-Authentication Denial-of-Service Loops in 4G and 5G}

\titlerunning{RRC Ping-Pong: Pre-Authentication Denial-of-Service Loops in 4G and 5G}

\author{
Elias Nechwatal\inst{2}$^{*}$
\and
Dominik Brunke\inst{1}$^{*\dagger}$
\and  
Nicolai Kröger \inst{2} 
\and 
Wolfgang Kellerer \inst{2} 
\and 
Michael Meier \inst{3}\inst{4}
\and
Michael Rademacher \inst{4}
\and
Hartmut König \inst{1}}
\authorrunning{E. Nechwatal, D. Brunke et al.}

\institute{ZITiS, Munich, Germany \and   Technical University of Munich, Munich, Germany \and University of Bonn, Bonn, Germany \and Fraunhofer FKIE, Bonn, Germany\\ \email{dominik.brunke@zitis.bund.de}}

\maketitle

\begingroup
\renewcommand{\thefootnote}{${*}$}
\footnotetext{The first two authors contributed equally to this work. It is based on the master's thesis by Elias Nechwatal "RRC-Based DoS via Coordinated Rogue Cells in 4G/5G Networks".}
\endgroup

\begingroup
\renewcommand{\thefootnote}{${\dagger}$}
\footnotetext{Corresponding author.}
\endgroup

\ifx\vordoi\empty\else
\begingroup
\renewcommand{\thefootnote}{}
\footnotetext{This preprint has not undergone peer review (when applicable) or any post-submission improvements or corrections. The Version of Record of this contribution is published in \vorvolume, and is available online at \url{https://doi.org/\vordoi}.}
\endgroup
\fi

\begin{abstract}
The availability of 4G and 5G networks underpins critical infrastructure, yet
the control plane's pre-authentication signaling remains a potent attack
surface. Prior work showed that a single \ac{fbs} can deny service by injecting
unauthenticated \ac{rrc} messages. Our pretests reveal that modern smartphones do not follow the specification uniformly, and
vendor-specific hardening such as retry budgets, \ac{ta}-scoped failure
counters, and signal-strength-driven reselection neutralizes most single \ac{bs}
techniques after only a handful of attempts.

We show that coordinated \acp{fbs} can circumvent these defenses. We introduce the \textit{Connection-Loop Attack}. The attack keeps the \ac{ue} cycling through pre-authentication setup on attacker-controlled cells --- via coordinated \acp{fbs}, or for the Reject variant a single cell --- so it never obtains the Idle reselection window needed to reach a legitimate cell.
Where the network bars a \ac{ue} by counting failures, the redirection also resets the attach-attempt counter, defeating the single \ac{bs} attack. Together these turn a time-limited nuisance into a persistent \ac{dos}.
From this concept we derive and
validate three concrete variants exploiting \textit{RRCRelease},
\textit{RRCReconfiguration}, and \textit{RRCReject} in 5G \ac{sa} and the
corresponding 4G messages, evaluating them on 11 modern smartphones spanning
Qualcomm, MediaTek, Apple, and Google Tensor Systems-on-Chip (SoC). The results
reveal pronounced chipset-dependent deviations and confirm loop viability in both
4G and 5G \ac{sa}, establishing a resource-efficient, signaling-based alternative
to interference-based \ac{dos}. We also discuss architectural countermeasures for
securing early signaling.

\end{abstract}

\section{Introduction}
\label{sec:intro}

\acresetall
4G and 5G networks have established mobile connectivity as a cornerstone of
critical infrastructure, powering everything from emergency services to
industrial \ac{iot}, so their availability is a safety requirement rather than a
convenience. Yet the mechanism that lets a \ac{ue} attach to a network also opens
a window in which the \ac{ue} must trust signaling it cannot yet authenticate,
notably \ac{rrc} control messages sent before \ac{as} security is activated.
Because these messages carry no integrity protection, an adversary operating a
\ac{fbs} can inject them to attack all three classical security goals: leaking
identifiers (confidentiality), tampering with configuration (integrity), and, our
focus here, denying service (availability) \cite{iso27001:2022}.

Historically, a single \ac{fbs} could deny 4G service by sending \textit{Attach Reject} messages with specific \ac{emm} cause
codes  or by injecting unauthenticated reject and release  messages \cite{jover2016lte,Shaik2016,yu2019emm}. These attacks were potent precisely because the \ac{ue} accepted the messages at face value.

As a starting point for this work, we re-ran the
classic single \ac{fbs} techniques against contemporary handsets with mixed results. Modern devices no longer follow the
specification uniformly, and vendor-specific hardening now truncates legacy
attacks after a handful of attempts. A single \ac{fbs} that spams
\textit{RRCConnectionReject} in 4G is abandoned by the \ac{ue} after roughly five
failures (see also \cite{Hussain2018LTEInspectorAS,Hussain5Gvuln}), which then attaches to the legitimate network. This hardening is the result of an
interplay between 3GPP specifications, operator certification requirements, and
vendor-specific interpretations.

Our central insight is that this can be defeated not by a stronger
message but by a better structure. The hardening mechanisms, retry budgets and
long-term barring timers, share a common assumption: that repeated failure is
confined to a single problematic cell or \ac{ta}, so that counting failures
there is a sufficient defense. We bring the formally described, theoretical Connection-Loop attack of Hussain et al.~\cite{Hussain5Gvuln} into practice and  extend it to 5G \ac{sa}, in which two coordinated \acp{fbs} in different
\acp{ta} redirect \acp{ue} back and forth using unauthenticated
pre-authentication signaling. 
In every variant \acp{ue} are denied the time in \ac{rrc} \textit{Idle} they need to scan for and reselect a legitimate cell. In the 4G case, where the network bars a \ac{ue} by counting failures, the redirection is treated as a legitimate mobility event and also resets the failure counter that would otherwise bar the \ac{ue}. A time-limited interruption thereby becomes a sustained \ac{dos}.
We used srsRAN \cite{srsRAN_Project_25_10} open-source stacks to test whether 5G's enhanced security architecture closes the loop, and find that in some respects it does not. The primary contributions of this work are as follows:
\begin{itemize}
    \item \textbf{Empirical characterization of modern UE hardening.} Pretests on
    contemporary smartphones characterize the retry budget and long-term-barring
    behavior that defeats legacy single \ac{fbs} \ac{dos}.
    \item \textbf{The Connection-Loop concept.} We implement and empirically validate a strategy for
    sustaining \ac{dos} by confining \acp{ue} to a state space that never
    returns to a free \ac{rrc} \textit{Idle} scan. We also identify the distinct
    deviation that each carrier message exploits. These are the counter reset
    scoped to the \ac{ta} for the 4G Release loop, as well as for the 5G Reconfiguration loop in combination with the acceptance of an
    unauthenticated handover, and backoff logic for the 5G Reject loop.
    \item \textbf{Three validated attack variants.} We implement and validate loop
    variants exploiting \textit{RRCConnectionRelease} (4G),
    \textit{RRCReconfiguration} (5G), and \textit{RRCReject} (5G), moving the
    formally-predicted-but-untested loop of Hussain et al.~\cite{Hussain5Gvuln}
    into practice and extending it to 5G \ac{sa}.
    \item \textbf{Cross-device evaluation.} We evaluate our attacks on 11 modern
    smartphones across Qualcomm, MediaTek, Apple, and Google Tensor
    \acp{soc}, revealing pronounced chipset-dependent variation in security
    enforcement.
\end{itemize}
The remainder is structured as follows. Section~\ref{sec:background} covers
\ac{rrc} pre-authentication signaling and the pretest evidence that motivates our
attack. Section~\ref{sec:related_work} situates the work in prior research.
Section~\ref{sec:threat_model} defines the threat model.
Section~\ref{sec:novel_attack_design} details the attack design and
Section~\ref{sec:experimental_setup} the testbed. Section~\ref{sec:evaluation}
presents the evaluation, Section~\ref{sec:countermeasures} discusses mitigations
and limitations, and Section~\ref{sec:conclusion} concludes this work.

\section{Background and Motivation}
\label{sec:background}

Our attack rests on a deviation between what the 3GPP specification  does not explicitly forbid during early
signaling and how devices behave. We first establish the specified
pre-authentication procedure and the control messages available to an adversary
(Section~\ref{sec:bg-preauth}), then present the pretest evidence that reveals the
exploitable deviation and motivates the Connection-Loop design
(Section~\ref{sec:bg-motivation}).

Both 4G and 5G \ac{sa} architectures consist of a \ac{ran} and a \ac{cn}. The
\ac{ran} handles the radio interface and the \ac{cn} manages authentication and
mobility~\cite{3gpp-36.331-r18,3gpp-24.301-r18,3gpp-38.331-r18}. We focus on the
\ac{rrc} layer, which governs the radio connection between the \ac{ue} and the
\ac{bs} (\ac{enb}/\ac{gnb}).

\subsection{The Pre-Authentication Window}
\label{sec:bg-preauth}

\begin{figure}[htbp]
\centering
  \includegraphics[width=0.85\linewidth]{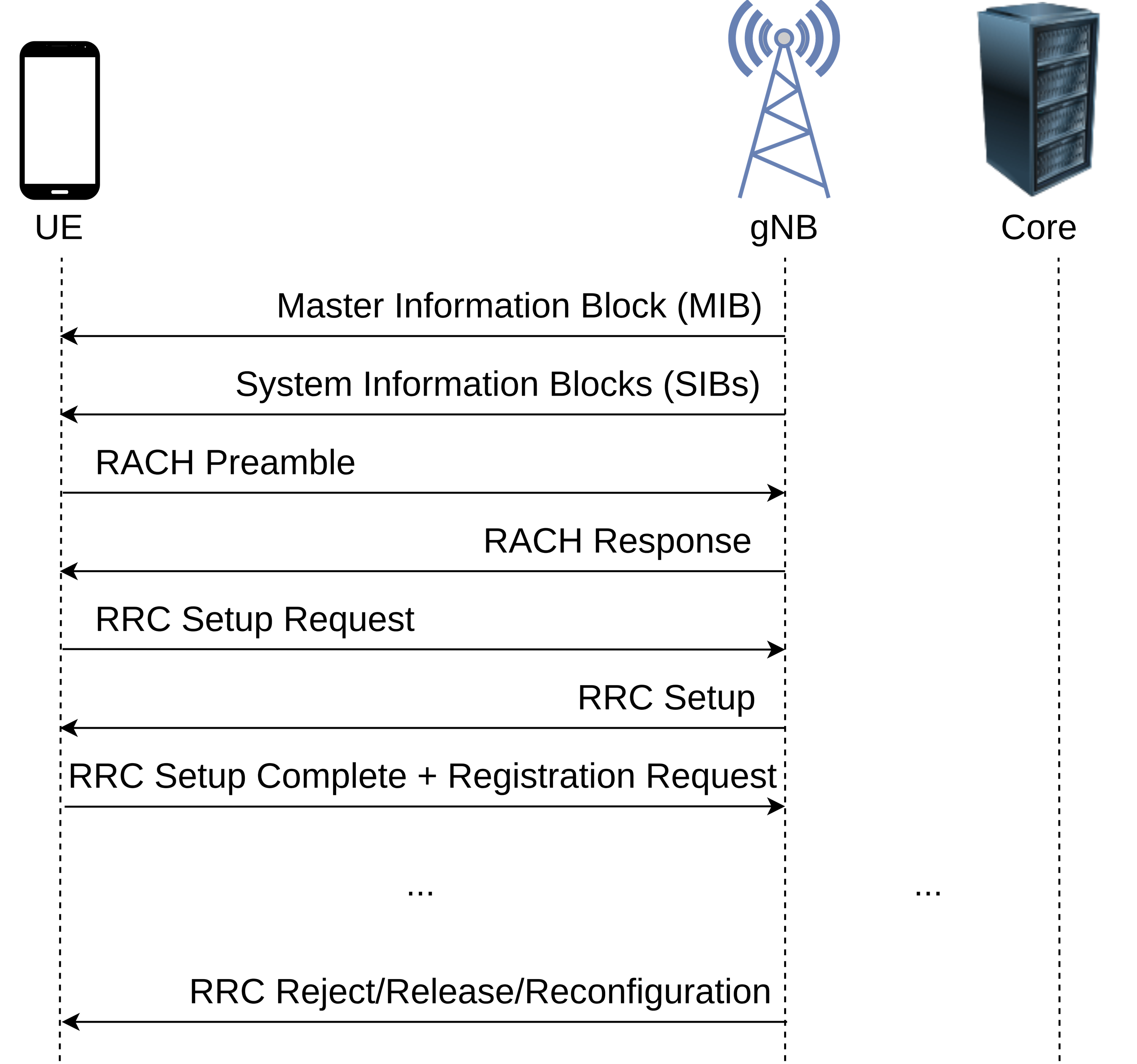}
\caption{Attach procedure in 5G, including \ac{mib}, \acp{sib}, \ac{prach} preamble, RRC, and NAS signaling \cite{3gpp-36.331-r18,3gpp-38.331-r18}.}\label{fig:protocol_flow}
\end{figure}

The establishment of an \ac{rrc} connection follows a canonical sequence. After
receiving \textit{\ac{mib}} and \textit{\acp{sib}}, broadcast messages responsible for the connection configuration, the \ac{ue} sends a \textit{RACH Preamble}. After receiving a \textit{RACH Response}, the \ac{ue}
sends a \textit{RRCSetupRequest}, the network responds with \textit{RRCSetup}
containing initial configurations, and the \ac{ue} confirms with
\textit{RRCSetupComplete}~\cite{3gpp-36.331-r18,3gpp-38.331-r18}. As illustrated
in Figure~\ref{fig:protocol_flow}, the \textit{RRCSetupComplete} message
typically carries the initial \ac{nas} payload, the \textit{NAS Registration Request}.

This entire sequence occurs before \ac{as} security is
activated. While the \ac{cn} authenticates the \ac{ue} and distributes
cryptographic keys, the initial \ac{rrc} signaling is unauthenticated and
not integrity-protected, creating a window in which a \ac{fbs} can inject control
messages to influence \ac{ue} behavior before the device can verify the network's
legitimacy. The same window facilitates attacks against all three security goals,
which in this case concerns availability. We analyze three classes of
\ac{rrc} messages an adversary can transmit during this window to disrupt
connectivity or steer mobility, the building blocks of our attack variants.

\textit{RRCRelease} is responsible for connection termination and redirection of the \ac{ue} from \textit{Connected} to \textit{Idle}. In both 4G and
5G, the \textit{redirectedCarrierInfo} field can force the \ac{ue} to a specific
target frequency. While 3GPP allows this field to be sent before authentication in 4G, it is explicitly forbidden in 5G SA. \cite{3gpp-36.331-r18,3gpp-38.331-r18}

\textit{RRCReject} is sent  when the network refuses a connection request (e.g., congestion). It typically
includes a \textit{waitTime} in 4G and 5G \ac{sa}, instructing the \ac{ue} to apply
a backoff timer before retrying. \cite{3gpp-36.331-r18,3gpp-38.331-r18}

\textit{RRCReconfiguration} modifies radio bearers and mobility parameters. Of particular interest is
\textit{mobilityControlInfo} (4G) or \textit{reconfigurationWithSync} (5G), which
triggers a handover. Although 3GPP mandates that handover commands be ignored
before security activation, any failure to enforce this would let an attacker
deterministically redirect a device. \cite{3gpp-36.331-r18,3gpp-38.331-r18}

\subsection{Pretests and Motivation}
\label{sec:bg-motivation}

The reason the messages above no longer suffice on their own lies not in the specification but in how modern \acp{ue} guard against repetition. To design a persistent attack we must first understand these guards, which we characterized through pretests on contemporary devices.

3GPP TS~24.301 defines attach-attempt counters and timer-based barring
(T3402)~\cite{3gpp-24.301-r18}, but remains unspecific about how \acp{ue} should
handle early \ac{rrc} failures and whether repeated rejections from a single cell
trigger the same barring logic as \ac{nas} failures. This ambiguity is filled
downstream by operator-specific certification rules. For instance, some operators
mandate that any \ac{rrc} failure, such as an \textit{RRCConnectionRelease} sent
before authentication, increment the attach-attempt
counter~\cite{verizon_lte_data_retry}. Once a threshold (e.g., five failures) is
reached, the \ac{ue} must observe the T3402 barring timer, blocking further
attempts for a prolonged period \cite{Hussain5Gvuln}.

Our experiments with contemporary devices (Table~\ref{tab:attack_devices_res})
confirm these operator requirements are enforced in practice. Five rapid connection
attach-attempts against a single cell drive the \ac{ue} into a \ac{dos} state lasting
10 to  12 minutes, consistent with the default 12-minute T3402 barring
timer specified in 3GPP TS~24.301~\cite{3gpp-24.301-r18} and independently
observed on commercial handsets during over-the-air testing~\cite{Park2022}. For an
attacker with a single \ac{fbs} this is only a temporary success, not a persistent \ac{dos}.

While probing this behavior, we use a counter-reset rule that is itself
specified but whose security assumption breaks under an active adversary. Per
3GPP TS~24.301~\cite{3gpp-24.301-r18}, a \ac{ue} resets its
attach-attempt counter when it enters a new \ac{ta}. We refer to the
resulting effect as \textit{\ac{tai}-based retry budgets}. The
specification frames this reset around legitimate mobility, so that a
user moving into a new area is not locked out. It carries no requirement that
the \ac{ta} change be authenticated. From a security standpoint it is a behavior
an adversary can use. If the attach-attempt counter resets on every \ac{tai}
change and the change can be induced by unauthenticated pre-authentication
signaling, an attacker who repeatedly moves the \ac{ue} across forged \ac{tai}
boundaries can reset the counter indefinitely.

This observation reframes the attacker's task against hardening that counts
failures. A single \ac{fbs} is defeated by the retry budget. Yet a coordinated
attack using multiple \acp{bs} in different \acp{tai} can keep the counter
perpetually reset \cite{Hussain2018LTEInspectorAS}. As Section~\ref{sec:evaluation}
shows, this counter reset is essential for the 4G Release loop, whereas the 5G
Reject and Reconfiguration variants are sustained by different deviations.
Forcing a \ac{ue} to switch \acp{tai} normally requires a
reselection process gated by signal-strength requirements and vendor-side
hardening, and any time spent in \ac{rrc} \textit{Idle} gives the \ac{ue} a
chance to find a legitimate cell. The challenge is therefore to move the \ac{ue}
between \acp{fbs} in different \acp{tai} using signaling alone, without
letting it enter \ac{rrc} \textit{Idle} long enough to trigger a legitimate
frequency scan. Meeting this challenge is what our Connection-Loop
concept (Section~\ref{sec:novel_attack_design}) is designed to do.

\section{Related Work}
\label{sec:related_work}

Attacks on cellular availability span the physical layer to the control plane. At
the physical end, Ludant et al.~\cite{ludant2021sigunder} demonstrated Sigunder,
which modifies arbitrary transmissions and disrupts connectivity across a limited geographic area~\cite{erni2022adaptover}. Such attacks require sustained transmission and physical proximity. We instead target the control plane, where a single
crafted message achieves a comparable effect at a fraction of the radio cost.

A first generation of control-plane attacks used individual unauthenticated
messages. Shaik et al.~\cite{Shaik2016} used crafted control-plane
messages to instruct a device to go offline while impersonating a legitimate base
station. Jover et al.~\cite{jover2016lte} showed that forged \textit{Attach Reject}
or \textit{\ac{tau} Reject} messages carrying specific \ac{emm} cause codes
(e.g., \#7 EPS services not allowed~\cite{Shaik2016}) could disable 4G
connectivity until reboot or SIM reinsertion. Yu et al.~\cite{yu2019emm} reproduced
and extended these attacks using srsLTE and real smartphones, confirming that 
devices accepted unauthenticated reject messages and suffered lasting denial or
downgrades. As our pretests (Section~\ref{sec:bg-motivation}) show, contemporary
devices have since hardened against these single-message techniques. This work addresses these limitations.

Hussain et al.~\cite{Hussain2018LTEInspectorAS} introduced LTEInspector, a formal
framework combining model checking and symbolic verification to uncover 4G
control-plane vulnerabilities. Extending this analysis to 5G~\cite{Hussain5Gvuln},
they identified the Uplink \ac{nas} Counter Desynchronization Attack and, most
relevant, an incarceration attack using unauthenticated \textit{RRCReject}
and \textit{RRCRelease} messages. By alternating between two \acp{fbs}, an
adversary redirects the \ac{ue} between frequencies and keeps it trapped in a
continuous connection-setup loop. This is, in essence, the Connection-Loop, but
Hussain et al. identified it purely through formal model checking and
never experimentally validated it, explicitly citing the lack of
available open-source 5G stacks at the time. Our work used this
formally-predicted but untested loop as its starting point. We implement it, quantify where it succeeds and fails across 11 current handsets, and identify two loop variants outside the original formal analysis: a 5G Reconfiguration loop using unauthenticated \textit{reconfigurationWithSync} acceptance, and a 5G Reject loop using permissive modem backoff. Where Hussain et al. proved the loop possible in a symbolic model, we establish which chipsets actually exhibit it and why.

A parallel line of work traps \acp{ue} in reselection loops. Purification et
al.~\cite{Purification2024} described an \ac{fbs} attack in which the adversary
establishes an \ac{rrc} connection and then withholds any response after the NAS
Registration Request. Fei \& Wang~\cite{fei2019lte} tested such an approach
successfully with an OnePlus~5, and Wu et al.~\cite{SiliWu2021} replicated it in a
hardware testbed (LG~K4 and an IoT device), observing \ac{imsi}
disclosure and prolonged loss of service. Subsequent
work~\cite{handover_vuln,blackhole} used the same technique or small variants to
trap \acp{ue} in reselection loops on commercial handsets, most recently on devices such as the Samsung Galaxy A53 5G, Motorola One 5G Ace, Xiaomi Mi9T,
and Apple iPhone~7~\cite{blackhole}. This entire class of attacks depends on the \ac{ue} returning to \ac{rrc} \textit{Idle} and performing reselection. The \textit{Idle} window is the mechanism they use. Our Connection-Loop inverts this. It sustains \ac{dos} by denying the \ac{ue} the \textit{Idle} phase altogether, so the two approaches are complementary rather than variants of one another.

Table~\ref{tab:fbs-attack-overview} summarizes the landscape of persistent
\ac{fbs}-based pre-NAS \ac{rrc} \ac{dos} attacks and their validation status. Two gaps stand out: the loop class was either formal-only or validated on handsets from 2015–2022, and no prior work validates it across vendors in both 4G and 5G SA. Our work closes both, validating the loop on 2024–2025 flagship SoCs across four chipset vendors.

\begin{table}[htbp]
\centering
\caption{Overview of persistent DoS attacks via \ac{fbs} pre-NAS RRC in 4G/5G environments}
\label{tab:fbs-attack-overview}
\setlength{\tabcolsep}{4pt}
\small
\begin{tabularx}{\linewidth}{|>{\raggedright\arraybackslash}p{3.4cm}|>{\raggedright\arraybackslash}X|l|}
\hline
\textbf{Attack} & \textbf{Tested Devices (Release year)}  &\textbf{Reference} \\ \hline\hline
Attach/TAU Reject with EMM Cause &
iPhone 6s Plus (2015), Huawei Honor 9 (2017), Meizu M5 Note (2016)  &\cite{jover2016lte,Shaik2016,yu2019emm}\\ \hline
Authentication Reject (Numb Attack) &
Not specified  &\cite{Hussain2018LTEInspectorAS} \\ \hline
Detach Attack &
Not specified  &\cite{Hussain2018LTEInspectorAS} \\ \hline
NAS Counter Desynchronization &
None (formal only)  &\cite{Hussain5Gvuln}\\ \hline
RRC Reject / Release Loop &
None (formal only)  &\cite{Hussain5Gvuln}\\ \hline
NAS Registration Request Replay-Hold &
None (simulated UE, srsRAN/Open5GS)  &\cite{Purification2024} \\ \hline
Idle Reselection Loop via Priority Manipulation &
OnePlus 5 (2017), LG K4 (2016), Samsung Galaxy A53 5G (2022), Moto One 5G Ace (2021),
Xiaomi Mi9T (2019), iPhone 7 (2016)  &\cite{handover_vuln,blackhole,fei2019lte,SiliWu2021} \\ \hline
\end{tabularx}
\end{table}

\section{Threat Model}
\label{sec:threat_model}

We adopt the classical \ac{fbs} threat model. The adversary is a local active
attacker who deploys a \ac{sdr}-based \ac{fbs} setup to disrupt mobile-service
availability for target \acp{ue} within a limited geographic area. The adversary
impersonates legitimate base stations and can broadcast forged \acp{sib},
manipulate cell-reselection priorities, offsets, and \ac{tac} values, and inject
any \ac{rrc} or \ac{nas} message that occurs prior to \ac{as} security activation,
including \textit{RRCReject}, \textit{RRCRelease}, and
\textit{RRCReconfiguration}. It can also present arbitrary Cell IDs, frequencies
(EARFCN/ARFCN), and \acp{plmn}, and, by rotating these identities across at least two coordinated \acp{fbs} to evade per-cell or per-frequency barring, the Connection-Loop turns this capability into a persistent attack. \cite{Dabrowski2014,Park2019,Shaik2016}

To isolate control-plane vulnerabilities from physical-layer attraction, we assume
the adversary has already attracted the target \ac{ue} to a \ac{fbs}, achievable
through established methods such as a stronger signal than legitimate cells,
overshadowing, or known handover vulnerabilities~\cite{handover_vuln,Shaik2016}. We treat
this initial attachment as a precondition and focus on the persistence of
the \ac{dos} state through pre-authentication signaling. Two constraints apply.
First, the adversary holds neither the \ac{sim} keys of the target \ac{ue} nor
access to the \ac{cn} (\ac{hss}/\ac{udm}), and so cannot decrypt traffic, forge
integrity-protected messages, or complete NAS authentication. Second, the attack
is bounded by the \ac{rf} range of the \acp{sdr}, the antenna configuration, and
proximity to the target, and it fails once the \ac{ue} establishes a secure
connection with a legitimate base station and the \ac{cn}.~\cite{Kotuliak2022,Purification2024}

\section{Design of the RRC Connection-Loop Attack}
\label{sec:novel_attack_design}

Section~\ref{sec:bg-motivation} established the weak point common to all modern
signaling-based \ac{dos}, namely the \ac{ue}'s time in a free \ac{rrc}
\textit{Idle} scan. This is the only phase in which the \ac{ue} can discover and
reselect a legitimate cell. The Connection-Loop removes this phase. Where the
hardening counts failures within a \ac{ta}, the loop also exploits the counter
reset triggered by a change of \ac{ta}. The concrete deviation used differs
per message.

\subsection{The Connection-Loop Concept}
\begin{figure}[htbp]
    \centering
    \includegraphics[width=0.65\linewidth]{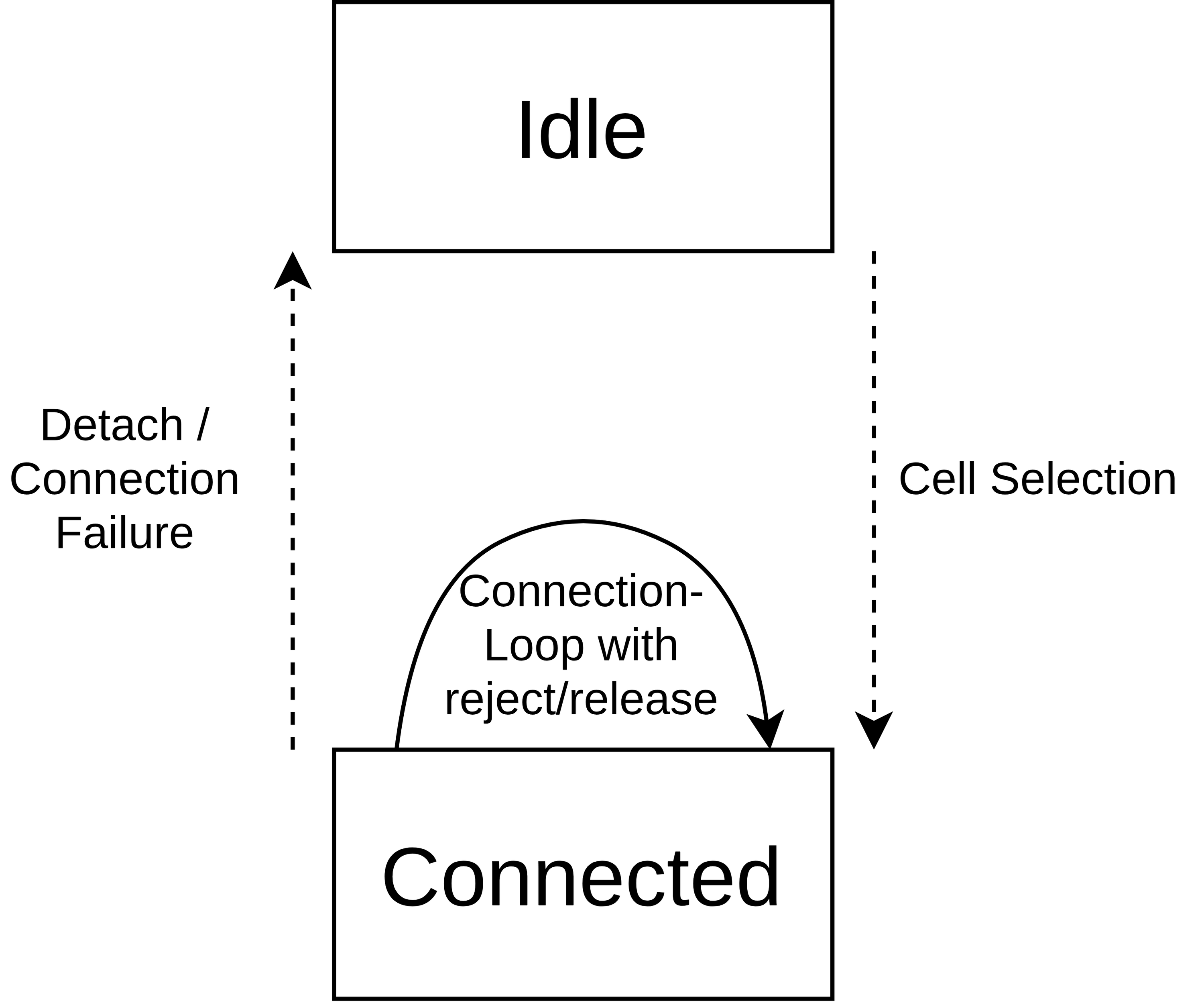}
    \caption{The Idle$\rightarrow$Connected flow represented with the dashed arrows and the Connection-Loop attack.}
    \label{fig:attack-loop}
\end{figure}

Figure~\ref{fig:attack-loop} depicts the state transitions of a \ac{ue} during a \ac{fbs} interaction. In the standard case (dashed arrows), the \ac{ue} moves from \textit{Idle} to \textit{Connected} following the cell selection and a detach, where, in the \textit{Idle} phase, the \ac{ue} may discover legitimate cells while scanning. Our strategy traps the \ac{ue} in a loop of immediate reconnections between \acp{fbs} so that the \ac{ue} never gets a free \ac{rrc} \textit{Idle} scan to trigger a legitimate reselection, see loop in  Figure~\ref{fig:attack-loop}. The loop can disrupt service for every UE within the coverage area of the \acp{fbs}, because any message of the loop is addressed to a specific UE. Its scope is constrained primarily by physical limitations and the computational resources of the \acp{fbs}.

\subsection{Attack Logic}
To sustain the \ac{ue} within the loop, the attack follows a four-step cycle:
\begin{enumerate}
    \item \textbf{Capture:} Attract the \ac{ue} to \ac{fbs}~A, e.g., via high signal
    strength.
    \item \textbf{Trigger:} Allow the \ac{rrc} connection to establish up to
    \textit{RRCSetupComplete}, but before \ac{as} security activation.
    \item \textbf{Redirect:} Terminate the connection with an
    unauthenticated \ac{rrc} message carrying a redirection command toward \ac{fbs}~B, located in a different \ac{tai}.
    \item \textbf{Loop:} \ac{fbs}~B repeats the process, redirecting the \ac{ue}
    back to \ac{fbs}~A.
\end{enumerate}
Using \textit{redirectedCarrierInfo} or an analogous steering field sent before
authentication in Step 3, the attacker turns a connection failure into a mobility
event initiated by the network, so the \ac{ue} does not perform a free
\textit{Idle} scan. Where the two \acp{fbs} advertise different \acp{tai}, the
\ac{ue} also treats the move as legitimate mobility and resets its retry budget.
This reset is required against hardening that counts failures
and not for the 5G Reject variant.

\subsection{Variant Taxonomy}
The loop can be instantiated through any pre-authentication message that
terminates the connection while steering mobility, without returning the \ac{ue} to
a free \textit{Idle} scan. This yields the three concrete variants we evaluate in Section~\ref{sec:evaluation}, each named for its carrier message:
\begin{itemize}
    \item \textbf{Release loop} (\textit{RRCConnectionRelease} with
    \textit{redirectedCarrierInfo}), a 4G vector exploiting redirection before security activation.
    \item \textbf{Reconfiguration loop} (\textit{RRCReconfiguration} with
    \textit{reconfigurationWithSync}), a 5G vector exploiting incomplete
    \ac{ue}-side enforcement of handover-before-security rules.
    \item \textbf{Reject loop} (\textit{RRCReject} with maximal \textit{waitTime}),
    a 5G vector exploiting permissive 5G retry logic.
\end{itemize}

\section{Experimental Setup}
\label{sec:experimental_setup}

\begin{figure}[htbp]
    \centering
    \includegraphics[width=0.6\linewidth]{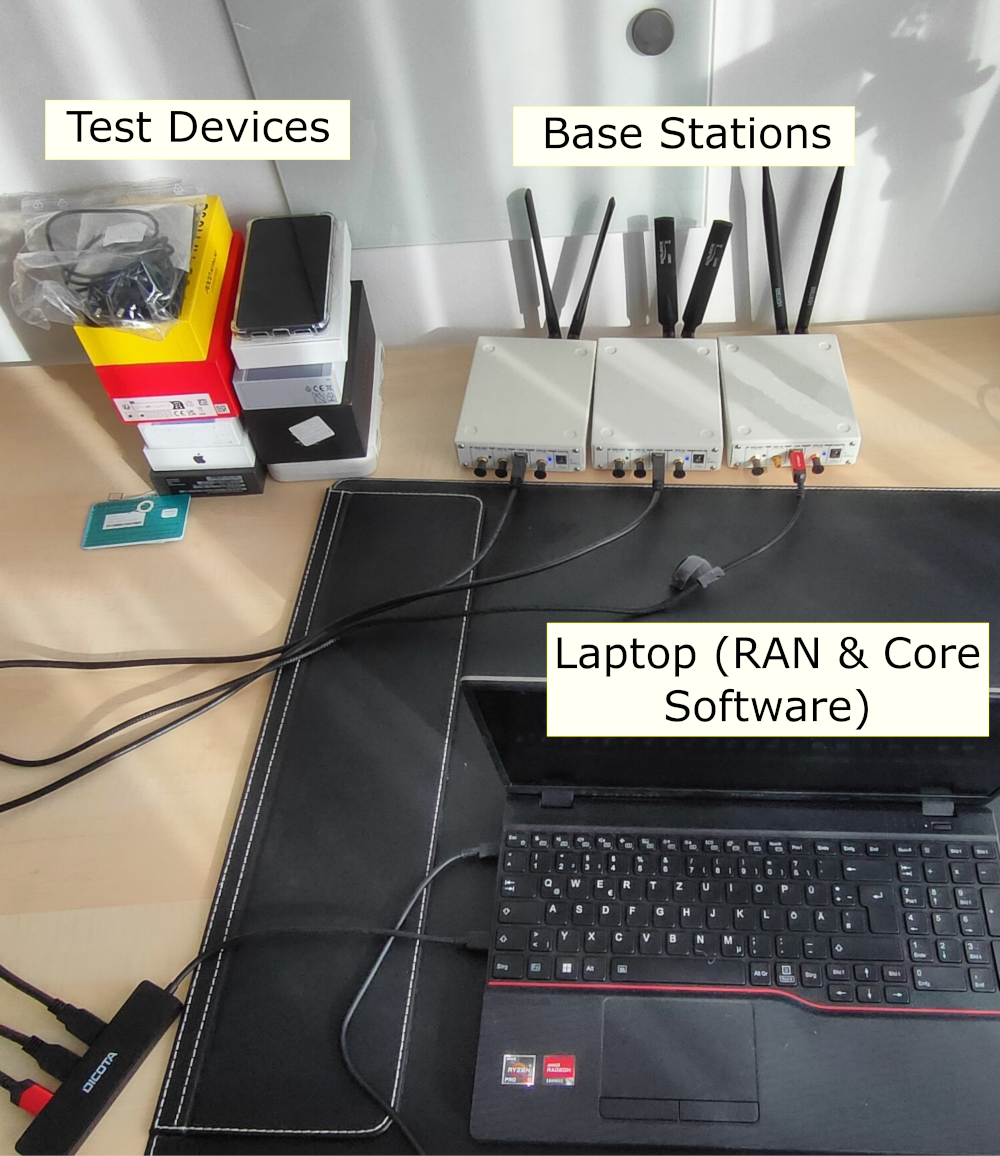}
    \caption{Setup utilized for Experiments.}
    \label{pic:setup}
\end{figure}

Our testbed is built around three Ettus \ac{usrp} B210 software-defined
radios~\cite{ettusb210}, two acting as \acp{fbs} and one as a legitimate \ac{bs}.
All connect to a single laptop (Fujitsu Lifebook E5512A, AMD Ryzen 5 Pro 5675U,
12 threads, 32~GB RAM) running a modified
srsRAN~\cite{srsRAN_4G_23_11,srsRAN_Project_25_10}. The testbed is orchestrated
with the open-source open5Gcube framework~\cite{open5Gcube_software,Horstmann2025open5Gcube},
which provides Docker-based automation to deploy srsRAN and Open5GS together.
open5Gcube originally supports only a single \ac{bs}, so we extended it to run
multiple parallel \acp{bs} with different Docker images and configurations. This is a
prerequisite for the loops which were part of the first author's master's thesis \cite{Nechwatal2026}. Programmable Sysmocom SIMs \cite{sysmocom_sim} emulated test and
real-world \acp{plmn}. Figure~\ref{pic:setup} shows the setup.

For the 4G attacks we used two \acp{fbs} running a modified srsRAN performing the attack and
 a third \ac{usrp} as a legitimate \ac{bs} for evaluation purposes built from the
unmodified official srsRAN 4G release. If a UE connects to the legitimate \ac{bs}, this indicates that no persistent DoS attack was achieved. For 5G \ac{sa} we extended the framework to
a multi-\ac{bs} environment, with open5Gcube's modular architecture enabling a
deployment analogous to the three-\ac{bs} 4G configuration. 

We combined \ac{ran}-side and device-side traces. Packet captures from the
modified srsRAN instance were recorded as pcap, and rooted Android phones were
instrumented with NetworkSignalGuru~\cite{networksignalguru} to log layer-3
messages. Experiment control logs, attacker state-machine traces, and hardware
timestamps were collected on the laptop for precise correlation between attacker
actions and \ac{ue} reactions. Target devices were \ac{cots} phones spanning
several vendors to capture vendor-dependent behavior
(Table~\ref{tab:attack_devices_res}).

\section{Experimental Evaluation}
\label{sec:evaluation}

We evaluate the Connection-Loop variants iteratively. First, we analyze individual
\ac{rrc} messages on a rooted reference device (Google Pixel~8 Pro) to establish
the signaling flow, then validate across the full set of commercial smartphones to
assess generality. Then we investigate whether unauthenticated 4G \ac{rrc} messages can trap a \ac{ue} in the loop. We then evaluate \textit{RRCConnectionReject} and \textit{RRCConnectionReconfiguration} followed by the Release loop with \textit{RRCConnectionRelease}. The extension of the analysis to 5G \ac{sa}  includes \textit{RRCRelease}, \textit{RRCReconfiguration}, and \textit{RRCReject} resulting in the Reconfiguration loop and the Reject loop.

\subsection{4G Attack Vectors}
We transmitted \textit{RRCConnectionReject} messages before and after
\textit{RRCConnectionSetupComplete}. The \ac{ue} consistently disconnected and
waited approximately 25 seconds before retrying. After five unsuccessful attempts,
however, the \ac{ue} abandoned the \ac{fbs} and connected to the legitimate
network. Manipulating fields such as \textit{waitTime}, \textit{extendedWaitTime},
or \textit{deprioritisationReq} produced no observable effect on the reconnection
interval or the number of attempts.

With \textit{RRCConnectionReconfiguration} we attempted a handover to a second \ac{fbs} via  \textit{mobilityControlInfo}. In all cases the \ac{ue} ignored the message and sent no
\textit{RRCConnectionReconfigurationComplete}, strictly compliant with 3GPP
TS~36.331~\cite{3gpp-36.331-r18}, which mandates rejecting mobility-related
reconfigurations before security activation.

\begin{figure}[htbp]
\centering
\includegraphics[width=\linewidth]{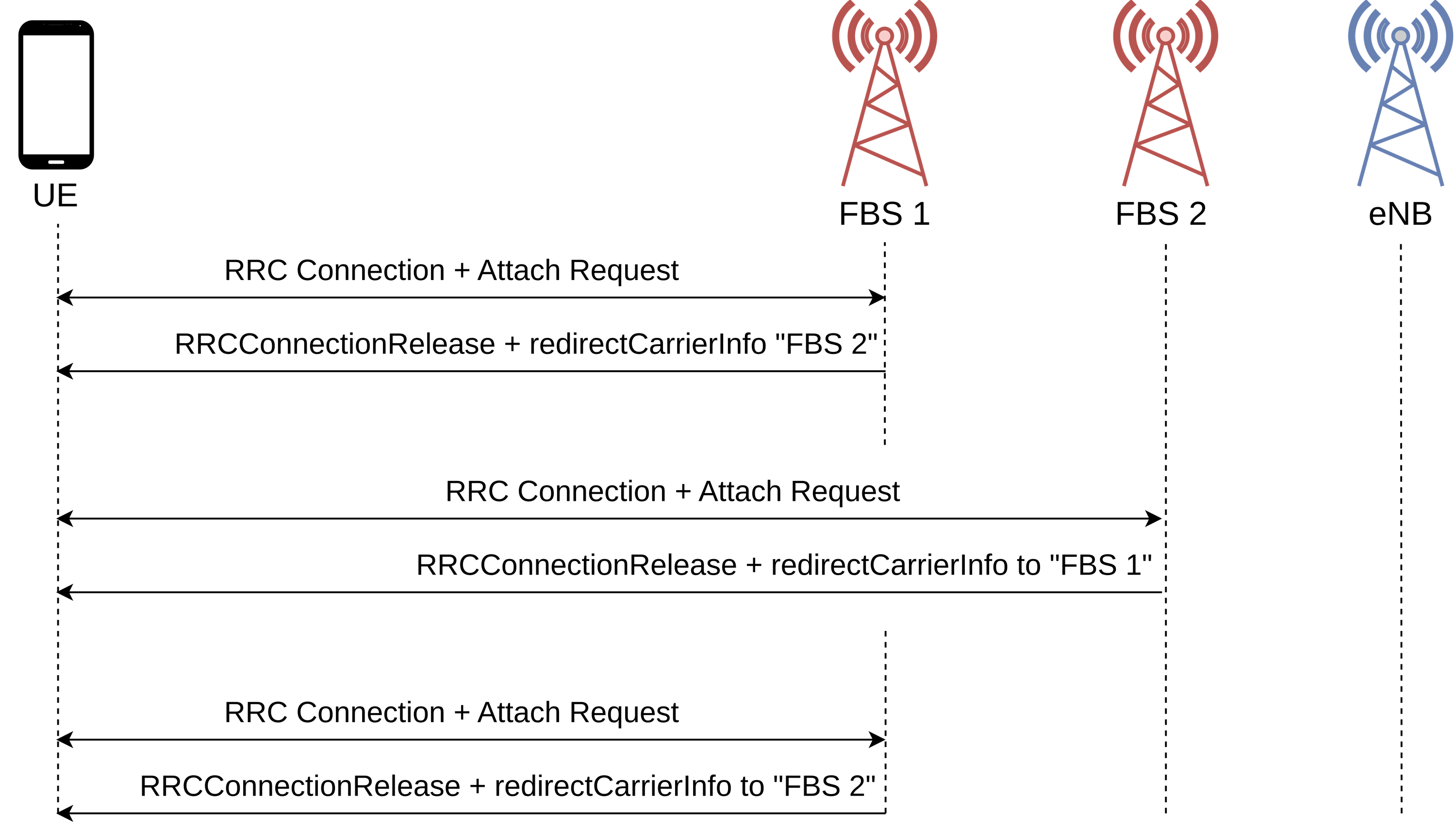}
\caption{4G Release loop flow.}\label{pic:release_attack_pcap}
\end{figure}

\paragraph{Release loop.}
\textit{RRCConnectionRelease} proved to be highly effective as the \textit{redirectedCarrierInfo} field triggered an immediate frequency switch.
Coordinating two \acp{fbs} to redirect the \ac{ue} toward each other established a
stable, bidirectional Connection-Loop (Figure~\ref{pic:release_attack_pcap}).  TS~36.331~\cite{3gpp-36.331-r18} does not forbid E-UTRA-to-E-UTRA redirection before security activation, letting the attacker deny the \ac{rrc} \textit{Idle} scan entirely and trap the \ac{ue}. The loop is driven by the redirection mechanism.

\subsection{5G Attack Vectors}
Attempting to reproduce the 4G Release loop with the 5G \textit{RRCRelease}
message, the \ac{ue} consistently ignored steering commands (across combinations of
\textit{cellReselectionPriorities} and \textit{redirectedCarrierInfo}) and reverted
to standard reselection. This resilience follows TS~38.331~\cite{3gpp-38.331-r18}, which mandates that \acp{ue} ignore all \textit{RRCRelease} fields except \textit{waitTime} when \ac{as} security is inactive.

\begin{figure}[htbp]
\centering
\includegraphics[width=\linewidth]{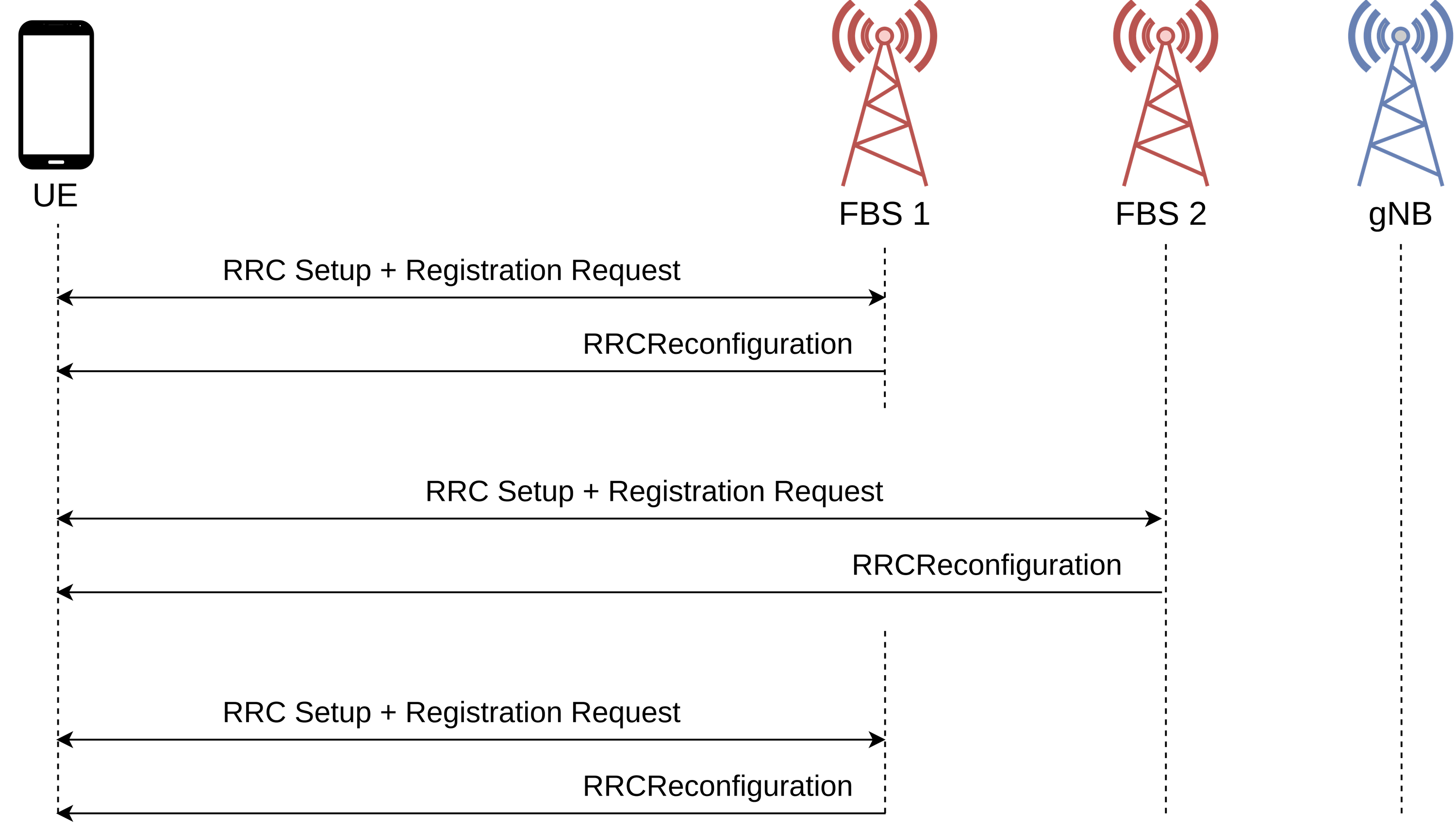}
\caption{5G Reconfiguration loop flow.}\label{pic:reconf_attack_pcap}
\end{figure}

\paragraph{Reconfiguration loop.}
In the 5G \textit{RRCReconfiguration} process using the
\textit{masterCellGroup} field with \textit{reconfigurationWithSync}, the \ac{ue}
accepted the handover command despite the absence of \ac{as} security. We achieved a sustained loop in intra-frequency setups
with different \acp{tac} (Figure~\ref{pic:reconf_attack_pcap}). TS~38.331~\cite{3gpp-38.331-r18} prohibits the network from sending unprotected handover commands but lacks a
corresponding requirement for the \ac{ue} to reject them. This asymmetry
allows a \ac{fbs} to force a handover. Notably, the \ac{ue} correctly
rejected \textit{radioBearerConfig} modifications, indicating inconsistent security
enforcement within the same message.

\begin{figure}[htbp]
    \centering
    \includegraphics[width=\linewidth]{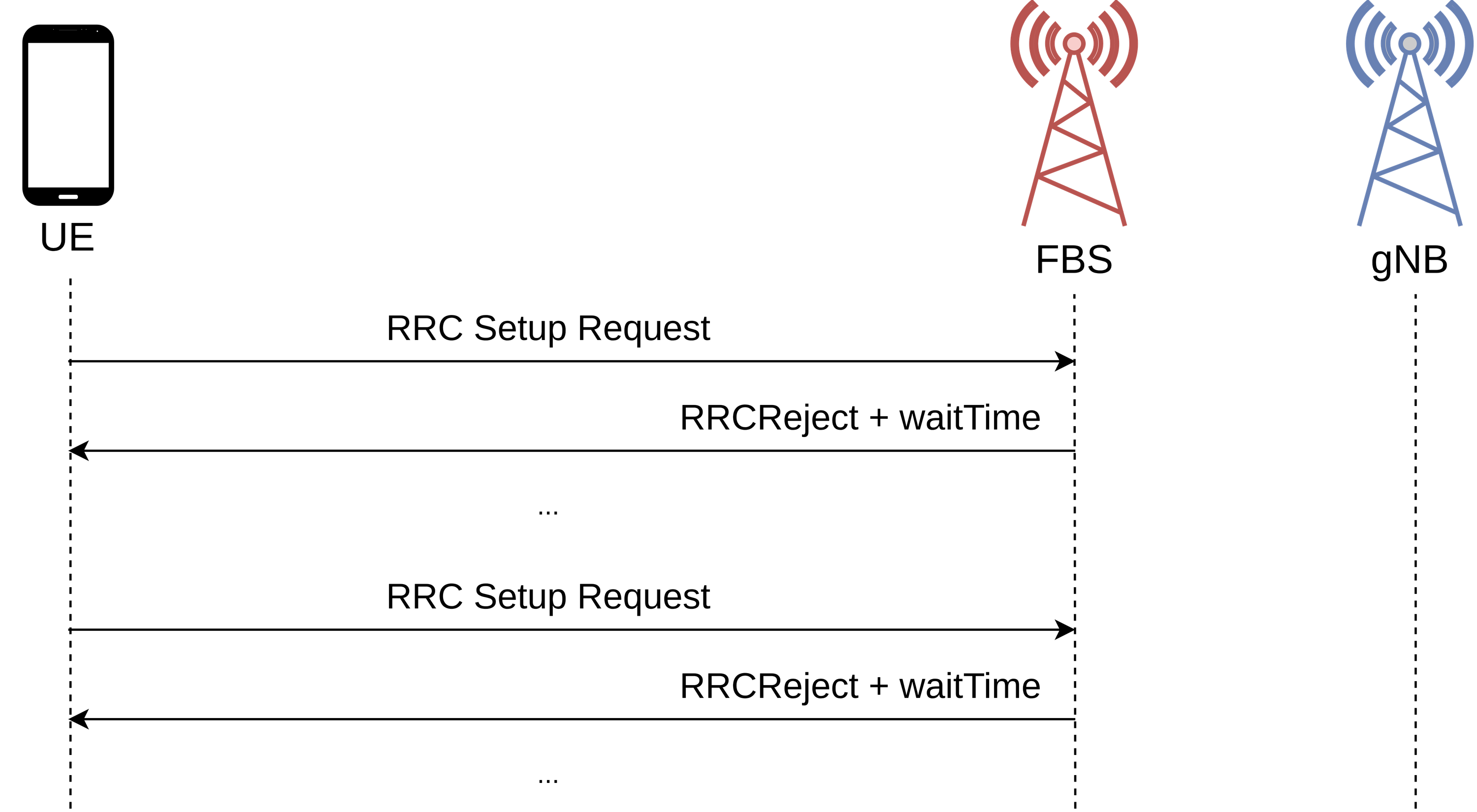}
    \caption{5G Reject loop flow.}
    \label{pic:rrc_reject_pcap}
\end{figure}

\paragraph{Reject loop.}
The \textit{RRCReject} vector, which failed in 4G, proved effective in 5G. Unlike
4G devices, the tested 5G \acp{ue} retried connection establishment without a
significant backoff, yielding a continuous loop. With the maximum
\textit{waitTime} (15~s), we maintained a \ac{dos} state for more than one hour using \acp{fbs} (Figure~\ref{pic:rrc_reject_pcap}). The 5G retry logic is thus more permissive than its 4G counterpart.

These results show that the 4G Release loop depends on the change of \ac{ta}. Two \acp{bs} in different \acp{ta} are
required to reset the attach-attempt counter, without which the \ac{ue} bars after
about five failures. The 5G Reject loop never approaches a barring threshold, because the tested modems apply negligible
backoff, so its persistence stems from permissive retry logic. The 5G Reconfiguration loop is enabled by the acceptance of an
unauthenticated handover and the change of \acp{ta}.

\subsection{Cross-Device Validation and Generalization}
To evaluate the practical impact of these deviations, we tested the identified
attack vectors on 11 commercial smartphones featuring various chipsets (Qualcomm, MediaTek, Apple, Google Tensor). The results are summarized in
Table~\ref{tab:attack_devices_res}.

\newcommand{\rothead}[1]{\rotatebox{90}{\textbf{#1}~}}
\begin{table}[htbp]
\centering
\caption{Overview of attack effectiveness across tested devices}
\label{tab:attack_devices_res}
\setlength{\tabcolsep}{3pt}
\small
\begin{tabularx}{\linewidth}{|l|l|>{\raggedright\arraybackslash}X|c|c|c|c|c|c|}
\hline
\textbf{Device Name} & \textbf{OS} & \textbf{SoC} &
\rothead{4G Release} & \rothead{4G Reject} & \rothead{4G Reconfig.} &
\rothead{5G Release} & \rothead{5G Reject} & \rothead{5G Reconfig.} \\
\hline\hline
ASUS ROG Phone 9      & Android 15      & Snapdragon 8 Elite            &\cmark &\xmark &\xmark &\xmark &\cmark &\xmark \\ \hline
Google Pixel 8a       & Android 15      & Google Tensor G3              &\cmark &\xmark &\xmark &\xmark &\cmark &\cmark \\ \hline
Google Pixel 8 Pro    & Android 15      & Google Tensor G3              &\cmark &\xmark &\xmark &\xmark &\cmark &\cmark \\ \hline
Google Pixel 10 Pro XL& Android 16      & Google Tensor G5              &\xmark &\xmark &\xmark &\xmark &\cmark &\crossedcmark \\ \hline
iPhone 16e            & iOS 18.3        & Apple A18                     &\cmark &\xmark &\xmark &\xmark &\cmark &\xmark \\ \hline
Nothing Phone 3a      & Android 15  & Snapdragon 7s Gen 3           &\cmark &\xmark &\xmark &\xmark &\cmark &\xmark \\ \hline
OnePlus 12R           & Android 14     & Snapdragon 8 Gen 2            &\cmark &\xmark &\xmark &\xmark &\cmark &\xmark \\ \hline
Realme GT             & Android 14    & Snapdragon 888                &\cmark &\xmark &\xmark &\xmark &\xmark &\xmark \\ \hline
Samsung Galaxy S25+   & Android 15      & Snapdragon 8 Elite            &\cmark &\xmark &\xmark &\xmark &\cmark &\xmark \\ \hline
Sony Xperia 5 V       & Android 14      & Snapdragon 8 Gen 2            &\cmark &\xmark &\xmark &\xmark &\cmark &\xmark \\ \hline
Xiaomi Poco M7 Pro 5G & Android 14 & MediaTek Dimensity 7025 Ultra &\cmark &\xmark &\xmark &\xmark &\cmark &\xmark \\ \hline
\multicolumn{9}{|c|}{\footnotesize \cmark~= attack succeeds \quad \xmark~= device resists \quad \crossedcmark~= partial success} \\ \hline
\end{tabularx}
\end{table}
The 4G Release loop was reliable across almost all devices. The notable
exception was the Google Pixel~10 Pro XL, which discontinued attempts after five
successful connections.

The 5G Reconfiguration loop with \textit{RRCReconfiguration} correlated strongly with
the SoC vendor. It was primarily effective on Google Tensor \acp{soc}, while all other devices adhered to the 3GPP requirements and ignored the unprotected reconfiguration. Pre-authentication security thus depends heavily on the modem vendor's implementation.

The 5G Reject loop with \textit{RRCReject} was the most consistent 5G vector, affecting almost
all devices. The only resilient one was the Realme GT, which
mirrored the 4G-style retry limit (5 attempts). The enabling deviation is thus
widespread across the latest modem generation.

\subsection{Controlled 4G-to-5G Transitions}
\label{sec:4gupgrade}
We investigated whether a \ac{ue} could be steered from a 4G \ac{fbs} to an attacker-controlled 5G \ac{fbs}. We evaluated four vectors: (i) 5G \ac{sa} redirection via \textit{RRCRelease}, (ii) overlapping EARFCN/ARFCN frequencies, (iii) pre-authentication \ac{endc} signaling, and (iv) inter-RAT handover via \textit{MobilityFromEUTRACommand}. None succeeded primarily due to strict adherence to 3GPP specifications~\cite{3gpp-36.331-r18,3gpp-38.331-r18}, which mandate ignoring NR-redirection and inter-RAT mobility commands unless \ac{as} security is active. The presented loops are limited to one \ac{rat}.

\section{Mitigation Strategies and Discussion}
\label{sec:countermeasures}

Our findings show that, despite 5G, the pre-authentication phase remains a critical
vulnerability. Low-cost \acp{sdr} and open-source stacks such as srsRAN simplify these attacks, shifting the threat from state-level actors to a broader range of adversaries.

The three validated loop variants differ in carrier message, but keep the \ac{ue}
within a \ac{dos} loop. By preventing a free \textit{Idle} reselection to a
legitimate \ac{bs}, the adversary bypasses the modem's internal hardening. The
hardening bypassed differs by variant. It is the retry budget scoped to the
\ac{ta} for the 4G Release loop, the retry and backoff limits for the 5G Reject
loop, and the combination of \ac{ta} retry budget and handover acceptance for the 5G Reconfiguration loop. A robust defense must address the structural
weaknesses to verify the origin of control-plane messages before they trigger
\ac{ue} state transitions.

\subsection{Proposed Countermeasures}
We propose a two-tiered defense, distinguishing an immediate heuristic fix from a
long-term architectural change.

A lightweight, implementation-level solution mandates a full transition to \textit{Idle} mode whenever multiple unauthenticated \ac{rrc} messages terminate or reset a connection
attempt. Currently, messages like \textit{RRCRelease} or \textit{RRCReject} can
trigger immediate actions such as redirection. Forcing a mandatory, full-spectrum
reselection based on actual signal strength stops
the loop at its defining step. This adds slight reconnection latency but forces the
adversary to re-attract the \ac{ue} through physical-layer dominance, substantially
increasing the attack's visibility to radio-monitoring systems. This complies with the current 3GPP specifications \cite{3gpp-38.331-r18,3gpp-38.304-r18,3gpp-36.304,Dabrowski2014,Park2019}. The Reject loop carries no redirection. It instead requires enforcing a minimum backoff or counting rejects across cells toward the barring threshold and could be mandated by  operator-specific certification rules like with \cite{verizon_lte_data_retry} in Section~\ref{sec:bg-motivation}.

An architectural solution requires cryptographic authenticity for early signaling. While
\textit{Broadcast but Verify}~\cite{ross2024fixing} addresses the static nature of
\acp{sib}, the \ac{rrc} messages require a flexible approach. The
\textit{\ac{hibs}} scheme of Yu et al.~\cite{HIBSRRCprotec} lets the \ac{ue} verify \ac{rrc}
and early \ac{nas} messages without a full public-key handshake. Integrating it into
the 3GPP standard prevents the Connection-Loop attacks demonstrated here, as the
\ac{ue} discards any \textit{RRCRelease} or \textit{RRCReconfiguration} not
signed by a legitimate network entity, at the cost of a non-trivial \ac{rrc}
state-machine update and increased signaling overhead and latency.

\subsection{Limitations}
This work has limitations. First, while we tested 11 diverse devices, the vast
ecosystem of modem implementations means some vendor-specific behaviors may remain
undiscovered. Second, our experiments used a controlled laboratory environment with
RF-shielded enclosures, and real-world multi-path fading and high-interference urban
environments may affect loop stability. Third, we focused exclusively on the
\ac{rrc} layer, leaving the \ac{nas} message injection out of scope.
More broadly, the tension between security and \ac{qos}
has historically favored the latter. The unprotected pre-authentication signaling was a design choice for seamless connectivity and attachment. The proliferation of \acp{sdr} and the shift toward \ac{oran} architectures necessitate re-evaluating this trust model.

\subsection{Ethics and Legal Notice}
All experiments were performed under strict safety and legal controls.
Transmissions were restricted to RF-shielded enclosures with minimal transmit
power to ensure that no external \acp{ue} or legitimate operators were interfered
with.

\section{Conclusion}
\label{sec:conclusion}

With modern smartphones no longer following
the specification uniformly and vendors hardening the pre-authentication phase
against legacy single \ac{fbs} \ac{dos}, signaling-based \ac{dos} is still possible. Our pretests characterized the hardening that defeats legacy single \ac{bs}
attacks and traced its shared weak point to the reselection window the \ac{ue} is
granted only in \ac{rrc} \textit{Idle}. Among the deviations that let an adversary
remove this window is a specified but security-critical assumption, namely that
the attach-attempt counter may reset on any tracking-area change.
Building on this, we implement and empirically validate the Connection-Loop, whose persistence rests on
denying the \ac{ue} a free \ac{rrc} \textit{Idle} reselection. Depending on the
carrier message, this is enabled, respectively, by the \ac{ta} counter reset (4G Release loop), the combination of \ac{ta} reset and unauthenticated-handover acceptance (5G Reconfiguration loop), and permissive retry logic (5G Reject loop). We validated the  predicted but untested loops in three concrete variants across 11 commercial devices. Susceptibility depends heavily on the modem chipset
and vendor, with some contemporary devices remaining unexpectedly vulnerable.

Our inter-RAT steering results show  that current devices
strictly enforce 4G-to-5G security boundaries, blocking unauthenticated upgrades. 
Fragmented, vendor-specific mitigations are insufficient against a determined
adversary. As open-source cellular stacks and low-cost \ac{sdr} hardware lower the
barrier to entry, the industry must move beyond security through obscurity, and as
we move toward 6G, a transition to a zero-trust control plane with authenticated
pre-authentication signaling is essential to the resilience of future cellular
generations.

\begin{credits}
\subsubsection{\discintname} The authors have no competing interests to declare that are relevant to the content of this article.
\end{credits}

\bibliographystyle{splncs04}
\bibliography{Bibliography}

\end{document}